\documentclass[twocolumn]{aastex63}

\usepackage{comment}

\usepackage{natbib}
\usepackage{multirow}
\usepackage{color}
\usepackage{bm}
\usepackage[normalem]{ulem}

\newcommand{\eg}{e.g., }

\def\gsim{\mathrel{\rlap{\lower 4pt \hbox{\hskip 1pt $\sim$}}\raise 1pt \hbox {$>$}}}
\def\lsim{\mathrel{\rlap{\lower 4pt \hbox{\hskip 1pt $\sim$}}\raise 1pt \hbox {$<$}}}

\shorttitle{Polarization of SN 2017 egm}
\shortauthors{S.Saito et al.}

\begin{document}

\title{ Late-Phase Spectropolarimetric Observations of Superluminous Supernova SN 2017egm \\
to Probe the Geometry of the Inner Ejecta }

\correspondingauthor{Sei Saito}
\email{s.saito@astr.tohoku.ac.jp}

\author{Sei Saito}
\affiliation{Astronomical Institute, Tohoku University, Aoba, Sendai 980-8578, Japan}

\author{Masaomi Tanaka}
\affiliation{Astronomical Institute, Tohoku University, Aoba, Sendai 980-8578, Japan}

\author{Takashi J. Moriya}
\affiliation{National Astronomical Observatory of Japan, National Institutes of Natural Sciences, 2-21-1 Osawa, Mitaka, Tokyo 181-8588, Japan}
\affiliation{School of Physics and Astronomy, Faculty of Science, Monash University, Clayton, Victoria 3800, Australia}

\author{Mattia Bulla}
\affiliation{Nordita, KTH Royal Institute of Technology and Stockholm University, Roslagstullsbacken 23, SE-106 91 Stockholm, Sweden}

\author{Giorgos Leloudas}
\affiliation{DTU Space, National Space Institute, Technical University of Denmark, Elektrovej 327, 2800 Kgs. Lyngby, Denmark}

\author{Cosimo Inserra}
\affiliation{School of Physics \& Astronomy, Cardiff University, Queens Buildings, The Parade, Cardiff, CF24 3AA, UK}

\author{Chien-Hsiu Lee}
\affiliation{NSF's National Optical-Infrared Astronomy Laboratory 950 N Cherry Avenue, Tucson, AZ 85719, USA}

\author{Koji S. Kawabata}
\affiliation{Hiroshima Astrophysical Science Center, Hiroshima University, Kagamiyama, Higashi-Hiroshima, Hiroshima 739-8526, Japan}
\affiliation{Department of Physical Science, Hiroshima University, Kagamiyama, Higashi-Hiroshima, Hiroshima 739-8526, Japan}

\author{Paolo Mazzali}
\affiliation{Liverpool John Moores University, Astrophysics Research Institute, IC2, 146 Brownlow Hill, Liverpool, Liverpool L3 5RF, Merseyside, United Kingdom}



\begin{abstract}
 
We present our spectropolarimetric observations of SN 2017egm, a Type I superluminous supernova (SLSN-I) in a nearby galaxy NGC 3191,
 with Subaru telescope at 185.0 days after the $g$-band maximum light.
 This is the first spectropolarimetric observation for SLSNe at late phases.
 We find that the degree of the polarization in the late phase significantly changes from that measured at the earlier phase.
The spectrum at the late phase shows a strong Ca emission line
and therefore we reliably estimate the interstellar polarization component assuming that the emission line is intrinsically unpolarized.
By subtracting the estimated interstellar polarization,
we find that the intrinsic polarization at the early phase is only $\sim$ 0.2  \%,
which indicates an almost spherical photosphere, with an axial ratio $\sim1.05$.
The intrinsic polarization at the late phase increases to $\sim$ 0.8  \%,
which corresponds to the photosphere with an axial ratio  $\sim1.2$.
A nearly constant position angle of the polarization suggests the inner ejecta are almost axisymmetric.
By these observations, we conclude that the inner ejecta are more aspherical than the outer ejecta.
This may suggest the presence of a central energy source producing aspherical inner ejecta.
 \end{abstract}



\newcommand{\ltsim}{\protect\raisebox{-0.8ex}{$\:\stackrel{\textstyle <}{\sim}\:$}} 
\newcommand{\gtsim}{\protect\raisebox{-0.8ex}{$\:\stackrel{\textstyle >}{\sim}\:$}} 

\section{Introduction}
\label{sec:intro}

Superluminous supernovae (SLSNe) are special types of supernovae (SNe), which are 10-100 times more luminous than normal SNe:
their luminosities reach \ltsim $-$21 mag \citep{Quimby2011, Chomiuk2011, DeCia2018, Lunnan2018}.
SLSNe are spectroscopically divided into two subclasses \citep{Gal-Yam2012,Moriya2018, Gal-Yam2019, Inserra2019}.
Type I SLSNe \citep[SLSNe-I, e.g.,][]{Quimby2011} do not show hydrogen  lines in their spectra
while Type II SLSNe \citep[SLSNe-II, e.g.,][]{Smith2010} show hydrogen  lines.

The origin of the extreme luminosities of SLSNe is not understood well.
Some candidates of power sources have been suggested: 
1) a large amount of $^{56}\mathrm{Ni}$ \citep{Wooslsy2007, Umeda2008, Gal-Yam2009, Nicholl2015a},
2) interaction between SN ejecta and circumstellar medium \citep[CSM;][]{Chevalier2011, Ginzburg2012, Moriya2013, Moriya2014, Chen2017a, Chandra2018}, and
3) additional energy injection from a central energy source such as a magnetar \citep{Kasen2010, Woosley2010, Mosta2014} or a black hole accretion disk \citep{Dexter2013, Moriya2018nov}.
However, the current observational data, mainly photometry and spectroscopy, do not necessarily distinguish these models in a conclusive way.

If a central energy source plays an important role,
the geometry of the explosion may become aspherical \citep{Mosta2014, Chen2016, Suzuki2017}.
Therefore, it is important to observationally probe the geometry of SLSNe.
SLSNe-I are in particular good targets to study the geometry of the inner ejecta thanks to the absence of the large hydrogen envelope.
Since we cannot spatially resolve extragalactic SNe, polarimetric observation is a powerful tool to study the geometry \citep{Wang2008}.

There are only four SLSNe-I observed with polarimetry:
SN 2015bn \citep{Inserra2016,Leloudas2017}, LSQ14mo \citep{Leloudas2015}, SN 2017egm \citep{Bose2018,Maund2019} and SN 2018hti \citep{Lee2019}.
Since the number of polarimetric observation is still small, a general picture of SLSNe-I has not been understood.
More samples are needed to probe the full 3D structure of the explosion shape of SLSNe-I.

The degree of polarization of SN 2015bn increases from $\sim$ 0.5 \% to $\sim$ 1 \% at $t \sim +20$ days
and it keeps $> 1 \%$ until $t \sim 45$ days
\citep[hereafter, $t$ represents the epoch from the optical maximum light in the rest frame]{Inserra2016,Leloudas2017}.
If the polarization is assumed to be intrinsic,
its ellipticity $E$ of the photosphere at $t = +$20 days is $E \sim 0.8$ \citep[its axial ratio $\sim1.2$;][]{Hoflich1991}.
In contrast, the degree of polarization of LSQ14mo from $t = -$7 days to $t = +$19 days is $\sim 0.2  \%$ and does not show a clear time evolution ($< 2\sigma$) during these epochs.
Therefore, the geometry of LSQ14mo is consistent with spherical symmetry \citep{Leloudas2015}.
For SN 2017egm, \citet{Bose2018} presented that SN 2017egm at early epochs ($t = -0.6, +5.2$ and $+9.1$ days) shows $\sim$ 0.5  \% polarization.
If the polarization is assumed to be intrinsic,
its ellipticity $E$ of the photosphere at the early epochs is $E \sim 0.9$ \citep[its axial ratio $\sim1.1$;][]{Hoflich1991}.
\citet{Maund2019} also measured the polarization of SN 2017egm at early epochs (from $t \sim 4$ days to $t \sim 19$ days),
which is broadly consistent with that measured by \citet{Bose2018}.
The degree of polarization of SN 2018hti keeps $\sim 1.9 \%$ from $t \sim -7$ days to $t \sim 14$ days,
but it is found to be comparable to the interstellar polarization,
suggesting that the geometry of SN 2018hti is almost spherical \citep{Lee2019}.
We note that no polarimetric observation of SLSNe-I at late epochs ($t > 50$ days) has been performed.

A difficulty in polarimetric observations is the effect of interstellar polarization (ISP), 
which consists of polarization components caused by dust extinction of both Milky Way and SNe host galaxies.
Since there is no straightforward way to reliably estimate the ISP, in particular that in the host galaxy, 
all the previous studies do not accurately estimate ISP;
some studies do not estimate ISP at all
and other studies \citep[including][]{Bose2018} estimate ISP from field stars,
from which only polarization caused by dust in our Galaxy can be estimated.

We here report our late-phase spectropolarimetric observations of SN 2017egm, which is a SLSN-I \citep{Nicholl2017}.
It was discovered by the $Gaia$ Satellite on 2017 May 23 \citep[MJD = 57896;][]{Delgado2017-05-25}
in a massive metal-rich spiral galaxy NGC 3191 \citep{Chen2017, Izzo2018, Bose2018} at redshift z = 0.0307 \citep{Dong2017}. 
By assuming Hubble flow,
the luminosity distance is $d_L = 136.6 \pm 1.0$ Mpc adopting the Hubble constant $H_0 = 67.4 \pm 0.5$ $ \rm{ km ~ s^{-1} Mpc^{-1} } $ \citep{Planck2018}.
SN 2017egm reaches maximum light in the $g$-band on 2017 June 21.8 (MJD = 57925.8) with the absolute magnitude of $M_{\rm g} = -20.97 \pm 0.05$ mag \citep{Bose2018},
which is comparable to the typical peak magnitude of SLSNe-I  \citep{Quimby2011, Chomiuk2011, DeCia2018, Lunnan2018}.
The light curves rise and decline linearly in magnitude from $t \sim -20$ days to $t \sim +20$ days
and do not show a pre-peak bump \citep{Bose2018}.
The spectra of SN 2017egm have O II absorption lines at $\sim$ 4100 \AA~ and $\sim$ 4400 \AA ~ in the early phases \citep{Bose2018},
which are the common features of SLSNe-I.
Thanks to its proximity, late-phase spectroscopic observations are also performed \citep[from $t=126$ days to $t=353$ days;][]{Nicholl2019}.

In this paper, we present the polarization spectra of SLSN-I SN 2017egm at a late phase and discuss the geometry of the explosion.
In Section \ref{sec:obs}, we describe our observations and data reduction.
In Section \ref{sec:res}, we show results of the observations.
In Section \ref{sec:dis}, we discuss the geometry of SN 2017egm suggested from the polarization data at the early and late epochs.
Finally, we give conclusions in Section \ref{sec:con}.

\section{Observations and Data Reduction}
\label{sec:obs}

\begin{deluxetable*}{cccccc}[t]
\tablewidth{0pt}
\tablecaption{Log of observations}
\tablehead{
  Object                 &           Date  &           Date  &           Exposure time &           Type\\
  & (UT) & (MJD) & (sec) & 
}
\startdata
SN 2017egm     &  2017 Dec 29.5 & 58116.5 & (600 × 4) × 6 & SLSN-I \\
HD 94851    & 2017 Dec 29.6 & 58116.6 & 20 × 4  & unpolarized std. \\
HD 251204  & 2017 Dec 29.4 & 58116.4 & 20 × 4  & polarized std.\\
Feige 34      & 2017 Dec 29.7 & 58116.7 & 20 × 4  & flux std. \\
\enddata
\tablecomments{  
  For all the targets, an offset slit with 0.8$^{\prime\prime}$ width, a 300 lines mm$^{-1}$ grism and Y47 filter were used, 
  covering 4400-9000 \AA ~with a resolution of $R = \lambda / \Delta\lambda \sim 600$.
  \\
}
\label{tab:log}
\end{deluxetable*}

Our spectropolarimetric observations of SN 2017egm were performed with Faint Object Camera and Spectrograph \citep[FOCAS;][]{Kashikawa2002} on Subaru telescope on 2017 December 29.5 (MJD = 58116.5). 
This epoch corresponds to 185.0 days from the maximum light in the $g$-band in the rest frame.

For all the observations presented in this paper, an offset slit with 0.8$^{\prime\prime}$ width, a 300 lines mm$^{-1}$ grism and an order sorting filter Y47 were used.
With this configuration, we cover 4400-9000 \AA ~with a resolution of $R = \lambda / \Delta\lambda \sim 600$.
FOCAS equips a rotating half-wave plate and Wollaston prism for measuring linear polarization.
Wollaston prism divides the incident ray into ordinary and extraordinary lights.
One set of the observation consists of four exposures corresponding to the four
position angles of the half-wave plate at 0$^\circ$, 22.5$^\circ$, 45$^\circ$ and 67.5$^\circ$.

For SN 2017egm, the exposure time for each frame is 600 seconds,
that is, the total exposure time of each set is 2400 seconds (600 seconds $\times$ 4).
We repeated this sequence for six times.
Summary of the observations is given in Table \ref{tab:log}.

We reduce the data according to a standard procedure by using IRAF.
Then, we derive Stokes parameters for each set of observations and combine them by rejecting apparent noises.
In this paper, Stokes parameters $Q$ and $U$ are defined as a fractional form \citep{Landi2002}:
$Q = (I_0 - I_{90})/I, U = (I_{45}-I_{135})/I$,
where $I$ is the total flux and $I_{\phi}$ is the flux with a polarizing filter rotated $\phi$ degree from the reference axis.
The position angle is $\theta \equiv 0.5 \tan ^{-1} (U/Q)$.
The degree of polarization is described as $P \equiv \sqrt{Q^2+U^2}$.
Because $P$ is positively biased due to the definition,
we also use debiased polarization $P^{\prime} \equiv  \sqrt{Q^2+U^2 - {\sigma_P}^2}$ \citep[e.g., ][]{Serkowski1974},
where $\sigma_P$ is the error of $P$.

To cover the wide wavelength range, we used an offset slit. Therefore, the instrumental polarization is not completely negligible.
To correct the instrumental polarization component, we observed an unpolarized standard star HD 94851.
The instrumental polarization is estimated to be $Q \sim -0.4~\%$ and $U \sim 0.4~\%$,
consistent with the specification of Subaru/FOCAS\footnote{https://www.naoj.org/Observing/Instruments/FOCAS/pol/ calibration.html}.
We fitted this component with quadratic functions and subtracted from the other data.
The residuals from the fitting is $\sigma = $ 0.14 \% both in Stokes $Q$ and $U$.
The offset of the position angle from the reference axis on the celestial plane was calibrated by observing a strongly polarized star HD 251204 (147 deg in \citealt{Turnshek1990}).
We note that the estimates of the position angle of HD 251204 differ among the literature
(155 deg in \citealt{Hiltner1954}, 153.3 deg in \citealt{Weitenbeck1999}  and 151.6 deg in \citealt{Ogle1999})
with discrepancies of up to 8 degrees.
A dome flat-field data through a fully polarizing 
filter were taken to correct for the wavelength 
dependence of the position angle of the equivalent 
optical axis of the half-wave plate.
Flux calibration was performed by using a flux standard star Feige 34.

\section{Results}
\label{sec:res}

\begin{figure}[t]
 \begin{flushleft}
    \includegraphics[width=\linewidth]{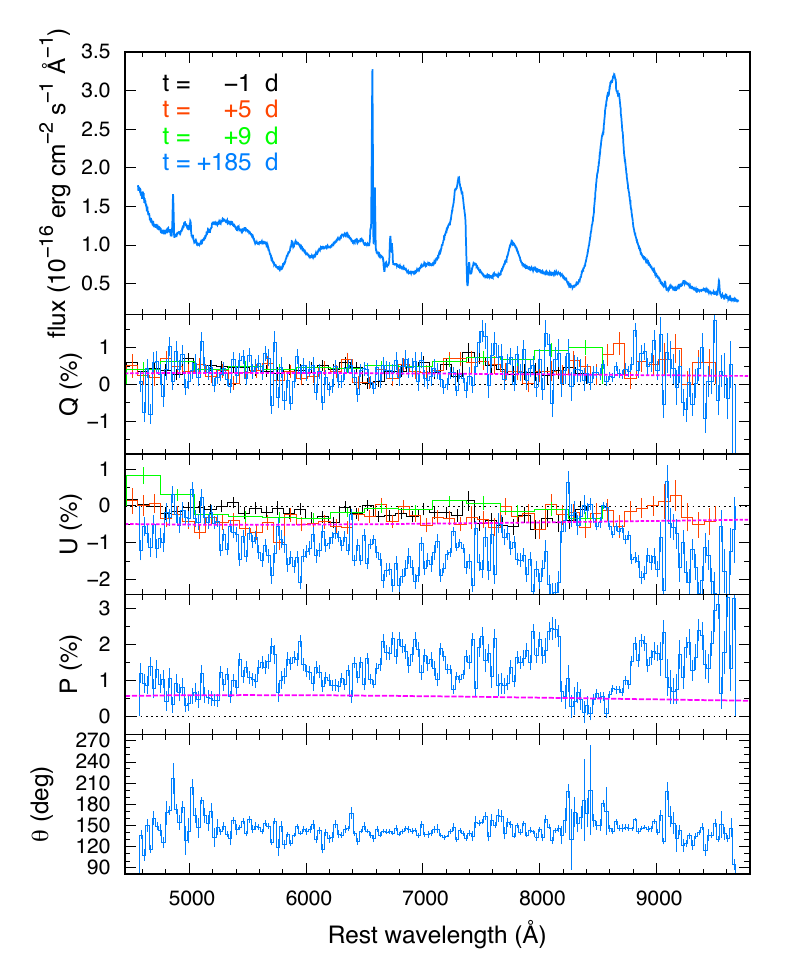}\\
\caption{
  \label{fig:spec}
  Top panel: Spectrum of SN 2017egm at $t = +185$ days. 
  Second, third and forth panels: observed Stokes parameters $Q$, $U$ and polarization $P$, respectively.
  The data at $t = +185$ days (blue lines) are binned to 30 \AA,
  while the data at $t = -1$ and  $+5$ days (black and orange-red lines, respectively) and at $t = +9$ days (green lines) are binned to 100 \AA ~and 300 \AA~\citep{Bose2018},
  respectively.
  Magenta dashed lines show the estimated ISP.
  Bottom panel: Position angle $\theta$ of SN 2017egm at $t = +185$ days.
  }
\end{flushleft}
\end{figure}

\begin{figure*}[t]
\begin{center}
  \begin{tabular}{c}

    \begin{minipage}{0.5\hsize}
      \begin{center}
        \includegraphics[width=\linewidth]{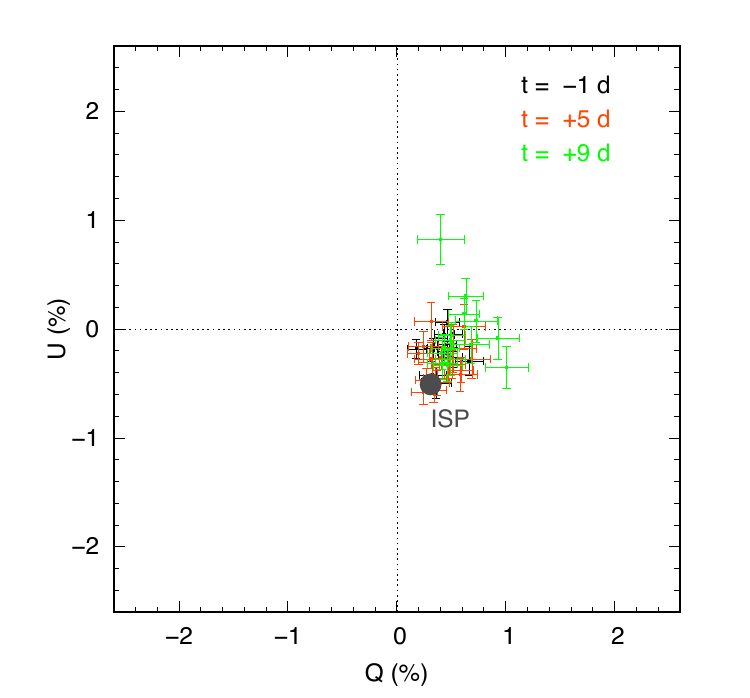}
      \end{center}
    \end{minipage}

     \begin{minipage}{0.5\hsize}
      \begin{center}
        \includegraphics[width=\linewidth]{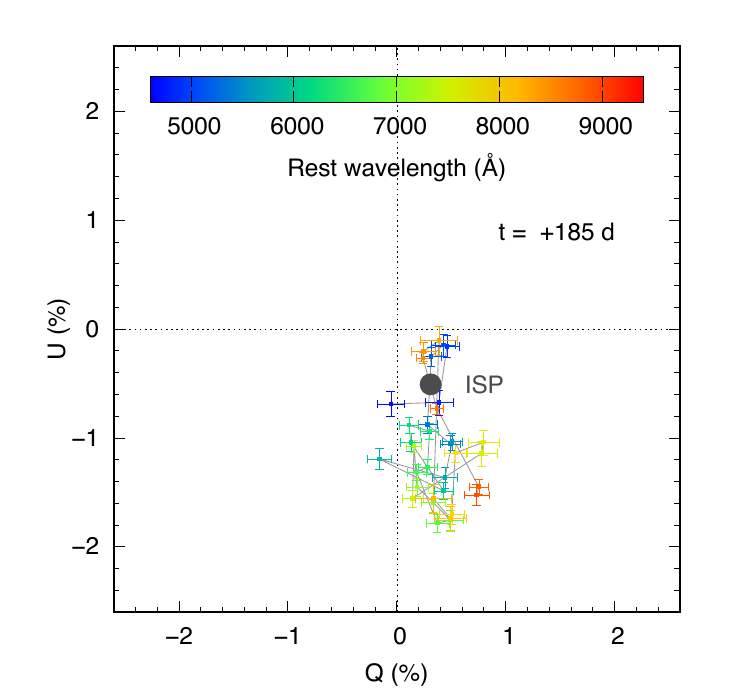}
      \end{center}
    \end{minipage}

  \end{tabular}
  \caption{Polarization of SN 2017egm in the $Q$-$U$ diagram. The gray circles in the both panels are estimated ISP at 8567 \AA.
  Left panel: The data at $t = -1, +5,$ and $+9$ days (black, orange-red and green points, respectively) binned to 300 \AA~\citep{Bose2018}.
  Right panel: The data at $t = +185$ days binned to 140 \AA~(note that this bin width is different from Figure \ref{fig:spec} and \ref{fig:int} for the visibility of the graph).
  The color of the points shows the wavelength in accordance with the colorbar.
  }
  \label{fig:qu}
  \end{center}
\end{figure*}

\begin{figure*}[t]
\begin{center}
  \begin{tabular}{c}
  
    \begin{minipage}{0.5\hsize}
      \begin{center}
        \includegraphics[width=0.95\linewidth]{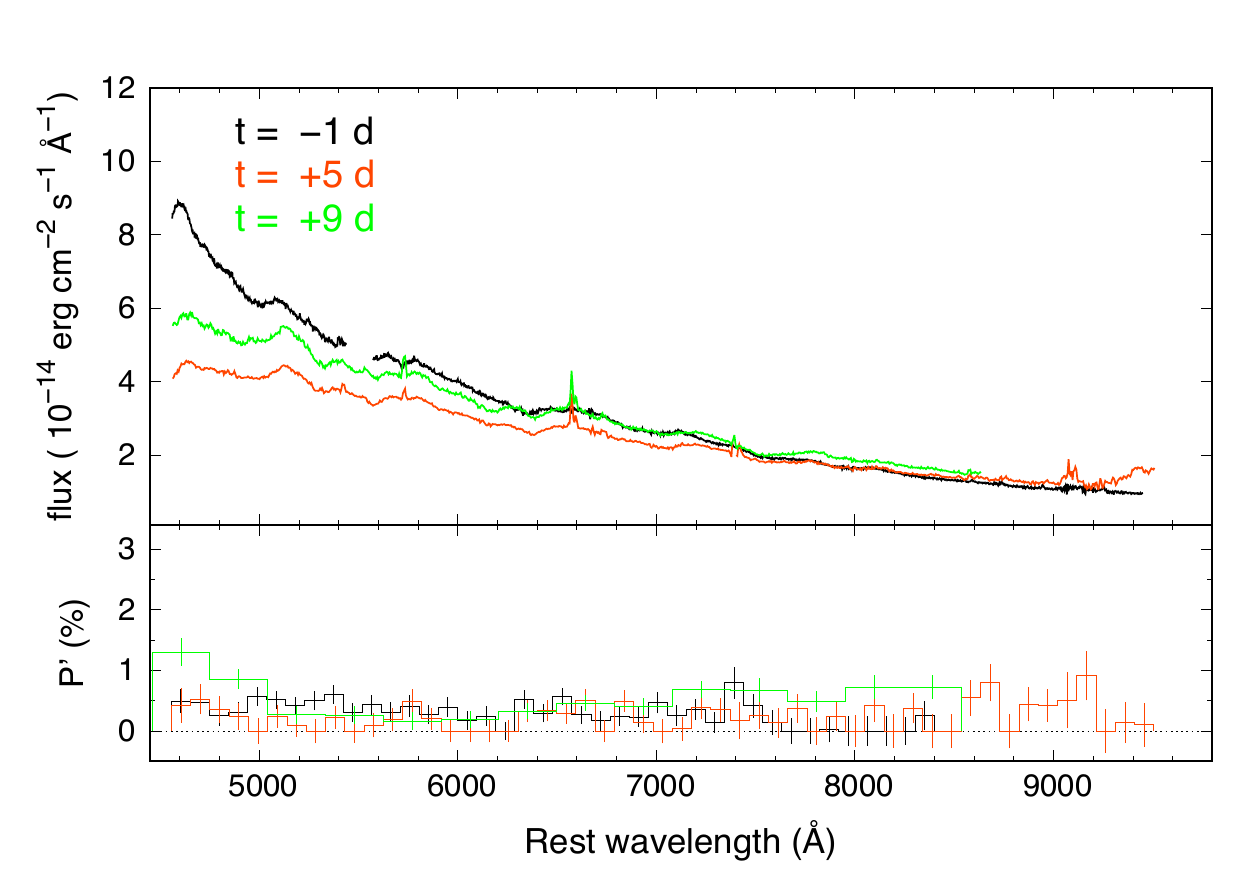}
      \end{center}
    \end{minipage}

     \begin{minipage}{0.5\hsize}
      \begin{center}
        \includegraphics[width=0.95\linewidth]{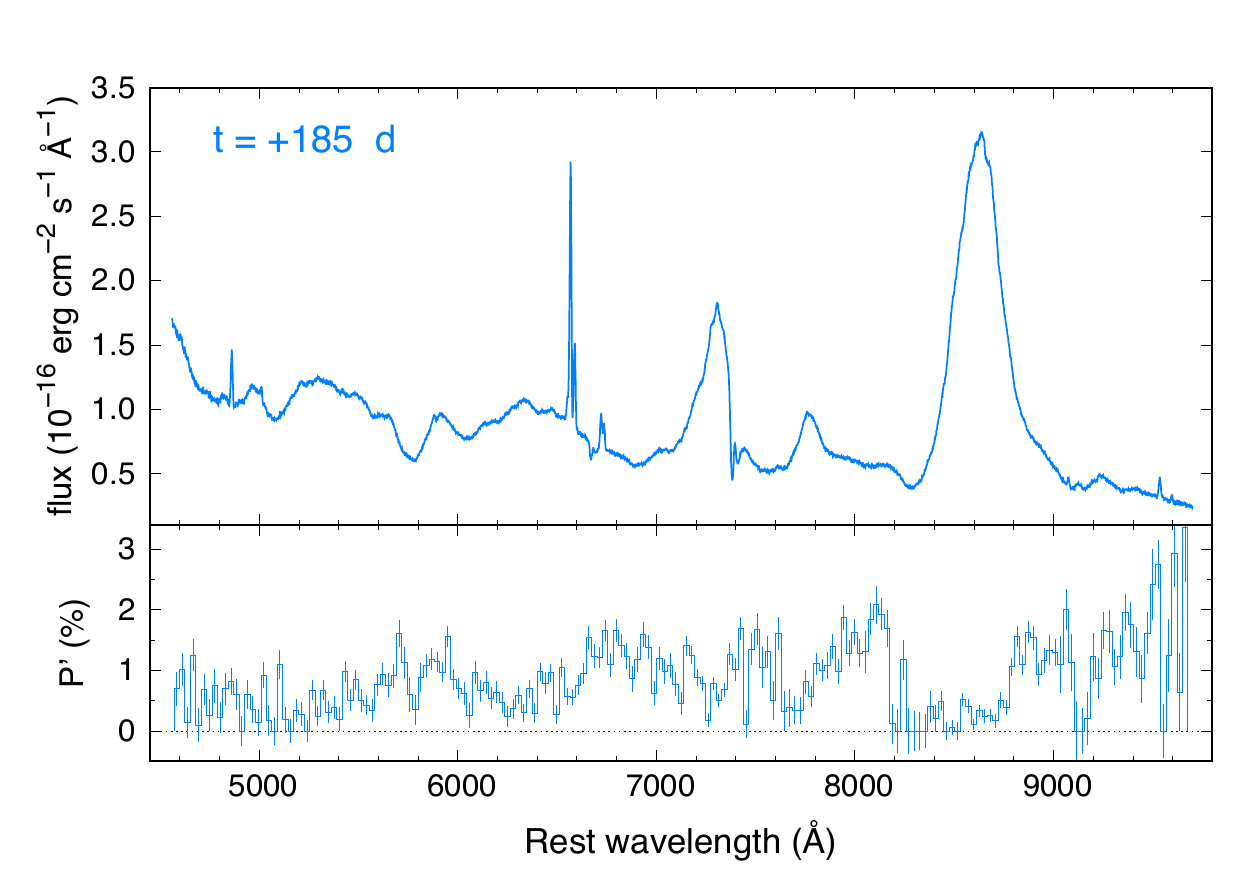}
      \end{center}
    \end{minipage}

  \end{tabular}
  \caption{Intrinsic polarization of SN 2017egm after subtraction of the estimated ISP.
  $P^\prime$ represents debiased polarization.
  Left panel: Spectra of SN 2017egm at $t = -1, +5$ and $+9$ days
  (black, orange-red and green lines, respectively; top panel) and ISP subtracted polarization (bottom panel).
  The polarization data at  $t = -1$ and $+5$ days binned to 100 \AA  ~and with the data at $t = +9$ days binned to 300 \AA~ \citep{Bose2018}.
  Right panel: Spectra of SN 2017egm at $t = +185$ days (top panel) and ISP subtracted polarization (bottom panel).
  The polarization data binned to 30 \AA.
  }
  \label{fig:int}
  \end{center}
\end{figure*}

In Figure \ref{fig:spec}, we show total flux, Stokes $Q$ and $U$, polarization and position angle of SN 2017egm at $t =185.0$ days.
Polarization data at $t = -0.6,~ 5.2,$ and $9.1$ days \citep{Bose2018} are also shown for comparison.
While the degree of $Q$ at the late epoch is nearly unchanged from the early epochs,
the degree of $U$ shows a significant time evolution.
The position angle at $t=185.0$ days is almost constant throughout all wavelengths.

Since the observed polarization consists of the intrinsic polarization of SN 2017egm and the ISP,
the ISP must be subtracted to study the geometry of the SN.
Here we estimate the ISP from the Ca II triplet emission lines around 8600 \AA.
Because the light radiated from atoms has no polarization
in principle \citep[e.g.,][]{Trammell1993} and the flux from the emission line
is mostly unscattered and unpolarized,
the degree of polarization at the wavelengths of lines can be regarded as the ISP.
In this way, as the SN itself can be used as an unpolarized reference, we can correct the sum of the ISPs from Milky Way and the host galaxy of the SN.
We take the wavelength range with flux of $>$ 2.0 $\times$ $10^{-16}$ erg cm$^{-2} $s$^{-1} $\AA$^{-1}$ 
as the Ca II triplet emission lines to minimize the contribution from the continuum light.
The fact that the degree of polarization is flat at the wavelengths of the Ca II triplet emission lines  supports that the emission lines are not only unpolarized but nearly completely unpolarized.
By averaging polarization degrees over this wavelength range,
ISP at the Ca II triplet lines is estimated to be $Q= 0.29 \pm 0.34 ~\%$ and $U= -0.46 \pm 0.38 ~\%$ including the uncertainties in the instrumental polarization.
Although the spectrum shows another prominent emission line around 7300 \AA~ ([Ca II]),
we did not use this line for the estimate of ISP because it is less prominent than the one around the Ca II triplet line and also affected by the telluric absorption.
For the wavelength dependence,
we apply Serkowski's law \citep[see Equation \ref{for:serkowski};][]{Serkowski1975}:
\begin{equation}
\label{for:serkowski}
P(\lambda) = P_{\rm{max}}\exp\left[-K \ln^2 (\lambda_{\rm{max}}/\lambda)\right].
\end{equation}
Here,  $\lambda_{\rm{max}}$ is the wavelength where ISP reaches the maximum value $P_{\rm{max}}$
and  $K$ is represented by $K = 0.01 + 1.66 \lambda_{\rm{max}}$ ($\mu$m) \citep{Whittet1992}.
We show the ISP in Figure \ref{fig:spec} by assuming  $\lambda_{\rm{max}}$ to be 5500 \AA,
which is a typical value for the ISP in the Milky Way.
It is noted that, although the properties of the dust in the host galaxy may be different,
we assume that the wavelength dependence of polarization in the host galaxy is the same as that in Milky Way.

Figure \ref{fig:qu} shows the comparison between the ISP and the observed polarization at the early and late epochs in the Stokes $Q$-$U$ plane.
As shown in Figure \ref{fig:spec}, Stokes $U$ show a clear time evolution. In the $Q$-$U$ plane, Stokes $U$ parameters in the early and the late phases are distributed in the opposite sides with respect to the ISP: the Stokes $U$ at the early phases shows positive $U$, while that at the late phase shows negative $U$ measured from the position of ISP.
The polarization at the late phase shows a larger deviation from the ISP (right panel) compared with the early phases (left panel).
The data points are localized in 
a small region in the $Q$-$U$ plane, suggesting that the position
angle measured from the ISP do not show a strong wavelength dependence at both the early and the late phases.

Figure \ref{fig:int} shows the degree of intrinsic polarization at the early (left panel) and the late (right panel) phases
after subtracting the estimated ISP.
Implications of the intrinsic polarization and its time evolution are discussed in Section \ref{sec:dis}.

\section{Discussion}
\label{sec:dis}

\subsection{Polarization at the early phase}

\citet{Bose2018} presented the polarization data of SN 2017egm at early phases.
They regarded the observed polarization to be intrinsic to the SN
since the wavelength dependence of the observed polarization is not
similar to Serkowski's law \citep{Serkowski1975}.
In this paper, we estimate the ISP in a reliable manner by using
the strong emission line at the late phase, providing more insight into the geometry of SN 2017egm.

We find that the degree of the observed polarization at the early phases ($t= -0.6, +5.2$ and $+9.1$ days)
is similar to the estimated ISP (see left panel of Figure \ref{fig:qu}).
Therefore, at these three epochs, the polarization intrinsic to the SN is quite low, $\sim$ 0.2   \% (left panel of Figure \ref{fig:int}), 
which corresponds to the ellipticity of the photosphere $E\sim$ 0.95 \citep[axial ratio $\sim1.05$;][]{Hoflich1991}.
The polarization is almost unchanged during these three epochs.
These facts suggest that the outer ejecta of SN 2017egm
\citep[$v \sim 10,000$ km s$^{-1}$ estimated from the absorption lines,][]{Bose2018} are almost spherical.

\subsection{Polarization at the late phase}

By subtracting the ISP from the late-phase data ($t=185.0$ days), 
we find that the intrinsic polarization increases from those at the early phases (right panel of Figure \ref{fig:int}).
Since the spectrum at this phase consists of many weak absorption and emission lines,
it is not straightforward to define the continuum polarization.
Therefore, we calculate a simple average over the wavelength range from 4450 \AA~to 8150 \AA.
The averaged intrinsic polarization at the late epoch is found to be $\sim$ 0.8  \%,
which corresponds to an ellipticity $E \sim$ 0.85 \citep[its axial ratio $\sim1.2$;][]{Hoflich1991}.
This implies that the inner ejecta of SN 2017egm ($v \sim 5,000$ km s$^{-1}$, estimated from Fe absorption lines) is considerably aspherical.
The ejecta have a nearly axisymmetric structure since the position angles measured from the ISP are almost constant as a function of the wavelengths as mentioned in Section \ref{sec:res}.

\begin{figure}[t]
 \begin{center}
    \includegraphics[width=\linewidth]{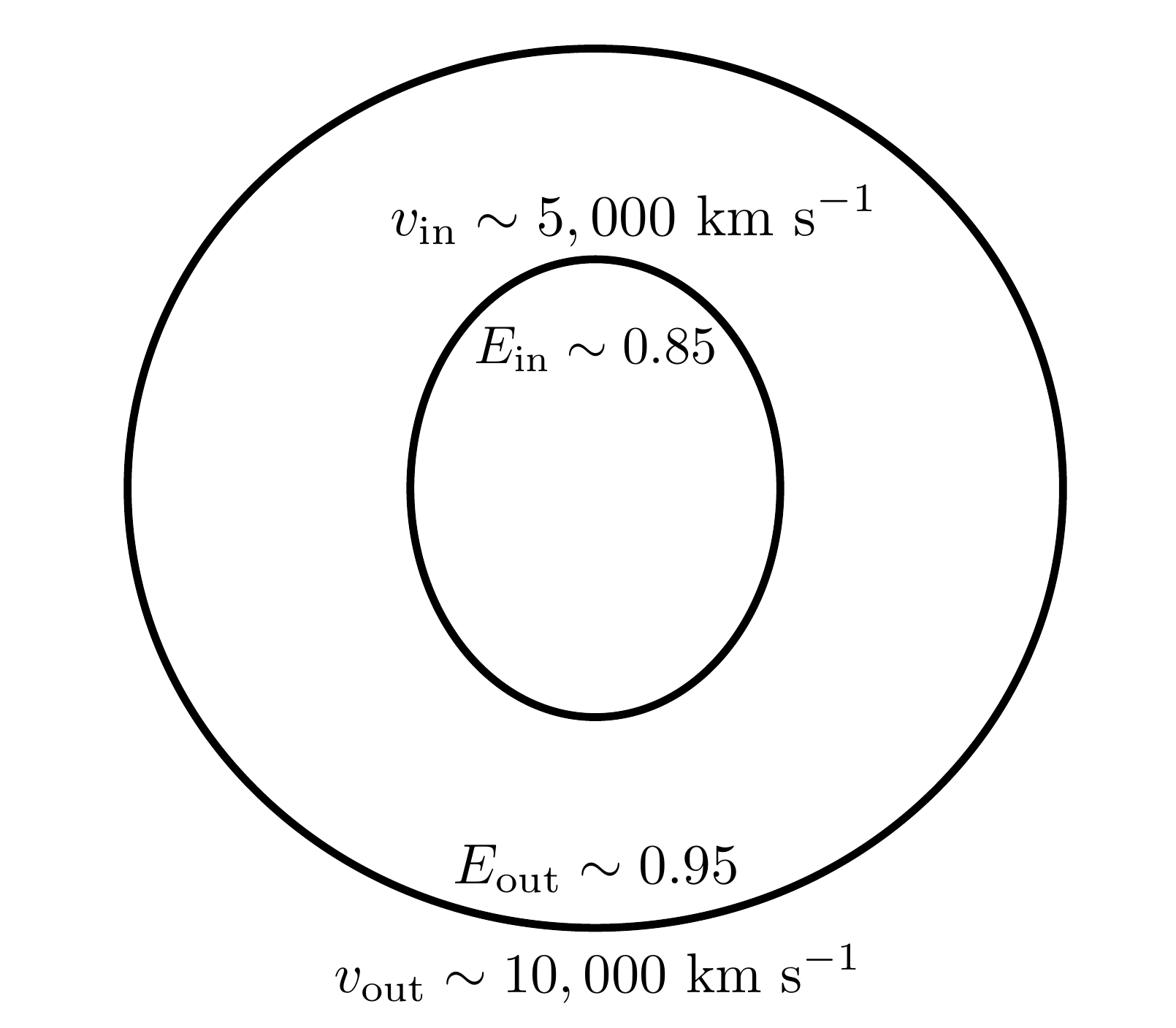}\\
    \caption{
  \label{fig:punch}
  A schematic picture of the ejecta geometry of SN 2017egm.
  The outer and inner ellipticities are estimated from the early- and late-phase data, respectively. The major axis of outer and inner ejecta are perpendicular to each other.
    }
\end{center}
\end{figure}

As noted in Section \ref{sec:res}, the observed Stokes parameters $U$ at the early phases are positive from the ISP (left panel of Figure \ref{fig:qu})
while those at the late phase are negative from the ISP (right panel of Figure \ref{fig:qu}).
In the $Q$-$U$ diagram,
the angle measured from the ISP changes by 180$^\circ$ from the early to the late phases.
In other words, the position angles on the sky changes by 90$^\circ$
since the angle measured in the $Q$-$U$ diagram corresponds to the position angle of $2\theta$.
This means that the direction of the SN major axis at the early phase is perpendicular to that at the late phase (see Figure \ref{fig:punch}).

The increase in the polarization at late phases has also been observed in Type II SNe \citep[e.g.,][]{Leonard2006,Chornock2010}. 
Such a behavior is also interpreted as an increase in the asphericity in the inner ejecta. 
There are, however, caveats on the interpretation. 
One is that the increase in polarization could also be caused by the decrease of electron-scattering opacity at the late phase even for a constant asphericity, as demonstrated by \citep{Dessart2011} for Type II SNe. Another is that 
the models by \citet{Hoflich1991} are calculated for early, optically thick phases. 
Although our late-phase spectrum of SN 2017egm still shows a clear continuum emission, 
the quantitative axis ratio may be subject to the uncertainty.

In addition to the continuum polarization, the polarization features across the absorption lines are of interest as they can be a probe of the elemental distribution \citep[\eg][]{Kasen2003,Tanaka2017}. 
In fact, our data show some changes in the polarization at the wavelengths of some absorption lines.
However, these parts of the data have low signal-to-noise ratios as they are absorption features, and thus,
we refrain from associating these features with the geometry of the ejecta.

\subsection{Implications}

We find that the intrinsic polarization of SN 2017egm increases from the early to the late phases.
SN 2017egm shows nearly zero polarization at the early phases,
while it shows a larger polarization at the late phase.
In other words, the outer ejecta \citep[$v \sim 10,000$ km s$^{-1}$; ][]{Bose2018} of SN 2017egm are almost spherical
while the inner ejecta  ($v \sim 5,000$ km s$^{-1}$) significantly deviate from spherical symmetry (see Figure \ref{fig:punch}).
The fact that the outer layer of SN 2017egm has no strong deviation from spherical symmetry is consistent with the findings by \citet{Coppejans2018},
who concludes that SN 2017egm does not have relativistic beaming of the radio emission.

It is interesting to note that the degree of polarization of SN 2015bn also increases from $\sim$ 0.5  \% to $\sim$ 1  \% 
at $t\sim$ 20 days and keeps $> 1 \%$ until $t \sim 45$ days.
SN 2015bn is also suggested to have an axisymmetric inner ejecta since the data points of SN 2015bn in the $Q$-$U$ plane are distributed along a straight line \citep{Inserra2016}.
These facts imply a similar configuration to SN 2017egm:
the outer ejecta of SLSNe-I are nearly spherical and the inner ejecta are more aspherical and these ejecta are nearly axisymmetric.
Although late-phase observations are not available for SN 2015bn,
this geometry might be general among SLSNe-I.

The ejecta geometry inferred from our observations may suggest the presence of a central energy source producing the aspherical inner ejecta.
In fact, a magnetar scenario \citep{Kasen2010, Woosley2010}, which is preferably discussed in the context of SLSNe,
may be able to produce aspherical ejecta \citep[e.g.,][]{Mosta2014, Suzuki2017}.
The same may also be true for the case of an accretion to a black hole \citep[e.g.,][]{Dexter2013, Moriya2018nov}.
It should be noted, however, that the exact impact to the ejecta geometry is not fully understood in both scenarios,
and thus, it is not straightforward to identify the origin of the aspherical inner ejecta.

It is also worth comparing the ejecta geometry from other probes.
In particular, line profiles in the late-phase spectra have been commonly used to study the element distribution.
In fact, for Type Ib/c SNe, aspherical element distributions are suggested by double-peaked or structured line profiles in the late-phase spectra \citep[e.g.,][]{Mazzali2005, Maeda2008, Modjaz2008, TanakaM2009, Taubenberger2009}.
In particular, a highly energetic SNe show a double-peaked profile with a wide separation, which may be a hint of a central energy source.
On the other hand, late-phase spectra of SLSNe-I typically show
smooth, single-peak profiles, which suggest nearly spherical inner ejecta \citep{Nicholl2019}.
Therefore, from the late-phase spectra,
SLSNe-I are suggested to have more spherical element distributions than normal stripped-envelope SNe.
This seems to be in conflict with our finding from the polarization as discussed above.
However, we should be careful in the comparison because the late-phase line profiles are sensitive to the element distribution
while the continuum polarization is sensitive to the density distribution (or more precisely, distribution of free electrons which produce polarization).

Since this is the first late-phase spectropolarimetric data for SLSNe-I,
it is too early to draw the firm conclusions about the general ejecta geometry of SLSNe-I.
Also, unfortunately, there are not many polarization data for normal stripped-envelope SNe \citep[\eg][]{TanakaM2012},
in particular those with a good time series except for a few cases \citep{Mauerhan2015, Stevance2017, Stevance2019}.
Therefore, it is not easy to readily compare the geometry between normal stripped-envelope SNe and SLSNe-I in terms of polarization.
In order to study the ejecta geometry and a central energy source of SLSNe-I in the context of SN explosions in general,
more time-series polarimetric observations are necessary both for normal SNe and SLSNe-I.

\section{Conclusions}
\label{sec:con}

We perform spectropolarimetric observations of SN 2017egm with FOCAS on Subaru telescope.
Thanks to the observations at the late epoch ($t=185.0$ days),
we reliably estimate the degree of the ISP from the strong Ca II triplet emission lines.
Then, using the estimated ISP,
we evaluate the intrinsic polarization of SN 2017egm at the early phases \citep[$t= -0.6, +5.2$ and $+9.1$ days; ][]{Bose2018} and at the late phase ($t=185.0$ days). 

We find that intrinsic polarization of SN 2017egm is $\sim 0.2  ~\%$  at the early phases (axis ratio of $\sim 1.05$)
while it increases to $\sim 0.8  ~\%$ at the late phase (axis ratio of $\sim 1.2$).
These facts indicate that the inner ejecta are more aspherical than the outer ejecta.
In addition, from the nearly constant position angles over the wavelengths, the inner ejecta are found to be axially symmetric.
Such a geometry might suggest the presence of a central energy source   producing a large asymmetry.

\acknowledgements

We thank the staff of Subaru Telescope for helping our observations.
We thank Subhash Bose and Jon Mauerhan for providing the polarization data of SN 2017egm.
We are grateful to the referee for careful reading of the manuscript and giving us useful comments.
This research was supported by the Grant-in-Aid for Scientific Research from JSPS (19H00694) and MEXT (17H06363).
G.L. is supported by a research grant (19054) from VILLUM FONDEN.


\end{document}